\documentclass[prl,twocolumn,showpacs,superscriptaddress]{revtex4}
\usepackage[dvips]{graphicx}
\usepackage{dcolumn}
\usepackage{bm}
\usepackage{amssymb}

\DeclareGraphicsExtensions{eps}

\begin{document}
\preprint{APS/123-QED}
\title{Direct evidence for a dynamical ground state in the highly frustrated Tb$_2$Sn$_2$O$_7$ pyrochlore}
\author{F. Bert}
\author{P. Mendels}
\author{A. Olariu}
\author{N. Blanchard}
\affiliation{%
Laboratoire de Physique des Solides, Universit\'{e} Paris-Sud,
B\^at.~510, 91405 Orsay, France}%
\author{G. Collin}
\affiliation{Laboratoire L\'eon Brillouin, DSM, CEA Saclay,
91191 Gif-sur-Yvette Cedex, France.}
\author{A. Amato}
\author{C. Baines}
\affiliation{Paul Scherrer Institut, Laboratory for Muon Spin Spectroscopy, CH-5232, Villigen PSI, Switzerland}
\author{A.~D. Hillier}
\affiliation{ISIS Facility, CLRC Rutherford Appleton Laboratory Chilton, Didcot, Oxon OX11 0QX, UK}

\date{\today}

\begin{abstract}
$\mu$SR experiments have been performed on powder sample of the
"ordered spin ice" Tb$_2$Sn$_2$O$_7$ pyrochlore compound. At base
temperature ($T=35$~mK) the muon relaxation is found to be of
dynamical nature which demonstrates that strong fluctuations persist
below the ferromagnetic transition ($T_C=0.87$~K). Hints of long
range order appear as oscillations of the muon polarization when an
external field is applied and also as a hysteretic behavior below
$T_C$. We propose a dynamical and strongly correlated scenario where
dynamics results from fluctuation of large spin clusters with the
"ordered spin ice" structure. 
\end{abstract}

\maketitle

In the field of highly frustrated magnetism, rare-earth (RE)
titanate pyrochlores RE$_2$Ti$_2$O$_7$ have attracted much attention
in the last decade. As a result of large spin values of the RE atoms
and usually small exchange coupling, spin anisotropy, dipolar and
exchange interactions compete. The delicate balance of these energy
scales, which varies with each particular pyrochlore compound,
stabilizes original magnetic phases from spin ice (RE=Ho, Dy) to a
unique collective paramagnetism in
Tb$_2$Ti$_2$O$_7$~\cite{Gardner99} still under debate. External
constraints such as magnetic field or pressure can easily
destabilize this balance and drive the system into complex phase
diagrams~\cite{Pentrenko04,Mirebeau04}. In the same trend of idea,
special attention was recently devoted to the closely related
stanate pyrochlore RE$_2$Sn$_2$O$_7$ compounds where lattice
expansion as well as modification of the RE oxygen environment can
also result in drastic changes of the balance of the interactions
and eventually in novel exotic ground states~\cite{Matsuhira02}.
Thus, while the titanate and stanate Tb pyrochlores exhibit similar
antiferromagnetic correlations at high temperature, the Ti compound
remains disordered and dynamical down to 70~mK whereas the stanate
counterpart undergoes a "ferromagnetic" transition at 0.87~K. It was
recently proposed from neutron experiments~\cite{Mirebeau05} that
the stanate compound freezes in an original uniform ($\mathbf{q}$=0)
spin ice structure where the four spins located at the vertices of
each tetrahedron obey the "two in two out" ice
rules~\cite{Snyder01}. The unexpected ferromagnetism would then
result from the alignment of the spin vector sums on each
tetrahedron. Interestingly, the Tb$^{3+}$ frozen moment deduced
independently from a nuclear Schottky anomaly in heat capacity
measurements is nearly twice smaller than the one deduced from
neutron scattering. Despite the static neutron picture, this
suggested the existence of slow fluctuations, out of the neutron
time window.

The coexistence of fluctuations, the fingerprint of frustration, and
glassy behavior~\cite{Uemura94,Dunsiger96,Bono04b,Bert05} appears as
a widespread and poorly understood feature of many frustrated
systems. Even more surprising is the recently reported coexistence
of spin dynamics and long range order, well below the transition
temperature of some pyrochlore compounds (RE=Gd,
Er)~\cite{Bertin02,Yaouanc05,Lago05}. Using the $\mu$SR technique,
we directly detect large spin fluctuations for the first time in the
Tb$_2$Sn$_2$O$_7$ compound which we attribute to fluctuations of
large spin clusters among the six-fold degenerate ground state of
the ordered spin ice structure.

Positive muon $\mu^{+}$ is a unique local probe to investigate
directly spin fluctuations. With a large gyromagnetic ratio
$\gamma_{\mu}=2\pi \times 135.5$~MHz/T and a weak coupling to its
magnetic surrounding, it is a very sensitive probe of magnetism. The
accessible time window usually falls in between that of NMR and
neutron experiments. As a noticeable example in the field of
frustrated magnets, $\mu$SR gave the first direct evidence of a
fluctuating ground state in the archetypal kagom\'e bilayers
SrCr$_{9p}$Ga$_{12-9p}$O$_{19}$~\cite{Uemura94}. In a $\mu$SR
experiment, the asymmetry of the $\mu^{+}$ decay between forward and
backward positron detectors is recorded as a function of the muon
life time in the sample. After subtraction of a background signal
arising from muons which miss the sample, the asymmetry is directly
proportional to the muon spin polarization $P(t)$. $P(t=0)$ equals 1 since the muon beam is 100$\%$ spin polarized.

$\mu$SR measurements on powder samples of Tb$_2$Sn$_2$O$_7$ in zero
and longitudinal applied field with respect to the muon initial
polarization were performed at ISIS and PSI facilities. The samples
were synthesized by standard solid state reaction and characterized
by X-Ray diffraction at room temperature and SQUID temperature
dependent susceptibility measurements.

\begin{figure}[!t]
\includegraphics[scale=0.8]{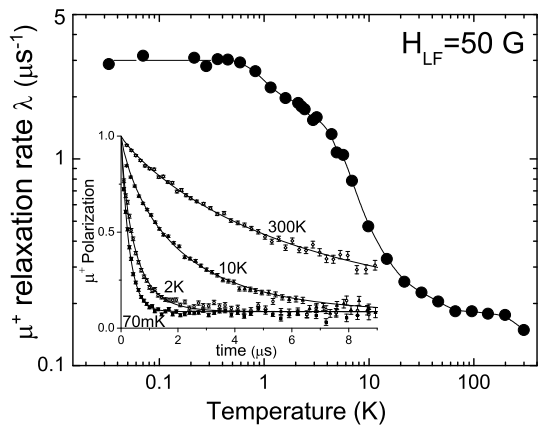}
\caption{\label{LF50} Variation with temperature of the muon
relaxation rate $\lambda$ in a small 50~G longitudinal field on a
$log-log$ scale. Inset: Muon polarization versus time at some
selected temperatures, also in 50~G.}
\end{figure}

The time dependence of the muon polarization has been recorded from
room temperature down to 30~mK in a small longitudinal field
$H_{LF}=50$~G. It could be fitted to a stretched exponential
function
\begin{equation}
\label{eqstretched} P(t)=e^{-(\lambda(T) t)^{\alpha}}+B
\end{equation}
with a stretched exponent $\alpha$ close to 1 in the whole
temperature range. The $T$ independent $B\simeq 10\%$ term stands
for the restored polarization of the muons which experience internal
fields, typically of nuclear origin, much smaller than $H_{LF}$ .
The muon relaxation rate $\lambda (T)$ is presented in
Fig.~\ref{LF50}. For a single time relaxation, it is expected to be
related to the electronic spin fluctuations rate $\nu$ by
\begin{equation}
\label{eqlambda} \lambda = \frac{2\gamma_{\mu}^{2} H_{\mu}^2
\nu}{\nu^2+\gamma_{\mu}^2 H_{LF}^2}
\end{equation}
where $H_{\mu}$ is the magnitude of the fluctuating field
experienced by the muon. For zero or small external field, we thus
get $\lambda \propto 1 / \nu$. While the muon relaxation rate hardly
depends on temperature at high temperatures, as a result of
paramagnetic fluctuations, it steeply increases below 10~K and down
to $\simeq 1$~K indicating a strong slowing down of spin
fluctuations at the approach of the ferromagnetic transition at
$T_C=0.87$~K. 
Below $T_C$ and down to the lowest temperature of the experiment
$T=30$~mK the muon relaxation rate saturates at a constant value.
 As depicted in the inset of Fig.~\ref{LF50}, there
is surprisingly no qualitative change in the shape of $P(t)$ above
and below $T_C$. In particular, the usual signs of (i) a static
ground state, namely the powder average long time tail $P(t
\rightarrow \infty, T \rightarrow 0)=1/3$, and (ii) long range
order, namely oscillations of the polarization due to well defined
internal field, are not observed.

\begin{figure}[!t]
\includegraphics[scale=0.8]{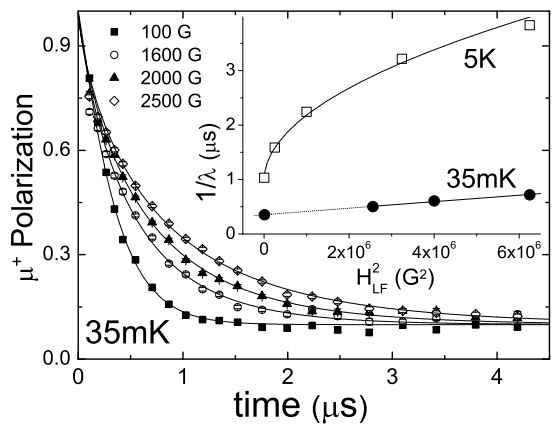}
\caption{\label{decouplage} Field dependence of the muon
polarization relaxation at 35~mK for large applied fields. Inset:
relaxation rates deduced from fits of the polarization with
Eq.~\ref{eqstretched} (solid lines in main figure) plotted against
$H_{LF}^2$ at 35~mK and for comparison at 5~K well above $T_C$. The
lines are linear fits of $1/\lambda$ versus $H_{LF}^2$ at 35~mK and
$H_{LF}$ at 5~K.}
\end{figure}

 In order to get more insight on the dynamical nature of the $\mu^+$
relaxation, we studied the magnetic field dependence $P(H_{LF},t)$
at base temperature. Were the relaxation at $T=35$~mK due to a
static local field $H_{\mu}$ at the muon site, or a distribution of
width $H_{\mu}$ of such static fields in a more disordered scenario,
$H_{\mu}$ would approximatively be given by $\lambda /
\gamma_{\mu}$. The full muon polarization $P(H_{LF},t)$ should then
be restored with $H_{LF} / H_{\mu} \gtrsim 5$. At 35~mK, $\lambda /
\gamma_{\mu} \simeq 35$~G and as shown in Fig.~\ref{decouplage} the
relaxation is still strong under 2500~G applied field. Therefore we
can safely conclude that the relaxation of the muon polarization is
of dynamical nature. For a dynamical relaxation, much higher fields
are needed than in the static case to suppress the $T_1$ processes.
The field dependence is then given by Eq.~\ref{eqlambda}. The field
dependence of the muon polarization in Tb$_2$Sn$_2$O$_7$ appears to
be rather complicated and we discuss it in more details in the
following. However, if one restricts to high field and long time
relaxation ($t \geqslant 0.1\mu$s) as in Fig.~\ref{decouplage},
$\lambda(H_{LF})$ nicely obeys Eq.~\ref{eqlambda} and we can extract
$H_\mu \simeq 20$~mT and $\nu \simeq 0.2$~GHz in a straightforward
manner. This fluctuation rate is below the accessible neutron
diffraction time window (typically in the GHz to THz range). At
higher temperatures, the $T_1=1/\lambda$ processes progressively
evolve to an unexpected linear dependence with $H_{LF}$ as shown in
the inset of Fig.~\ref{decouplage}. Such a linear field dependence
of $1/\lambda$ was formerly observed in
Tb$_2$Ti$_2$O$_7$~\cite{keren04,keren04b} and suggests similar
dynamics in both compounds. In the case of Tb$_2$Ti$_2$O$_7$, this
behavior was tentatively described by a power law decay of the spin
time correlation function.

We now address the second feature of the low T relaxation, namely
the absence of oscillations in zero external field. A basic scenario
to reconcile long range order and dynamics is to assume that the
spins fluctuate around the mean long range order in a magnon type
picture. The absence of oscillations would then arise from very
large and incoherent fluctuations so that the instantaneous
distribution of the magnitudes of the internal field at the muon
site is wide enough to strongly damp the oscillations (in the
extreme case of a fully disordered instantaneous field distribution
one retrieves the well known dynamical Kubo-Toyabe polarization
function which yields an exponential decay in the fast fluctuation
limit~\cite{hayano79}). Since the spin fluctuation rate $\nu$
 is much lower than the neutron time window, neutron
scattering experiments are sensitive to the instantaneous spin
structure. The well defined magnetic Bragg peaks which yield a
relatively large correlation length $l_C \simeq 18$~nm, together
with the full value of the Tb moment detected at low
T~\cite{Mirebeau05,gingras00}, invalidate here such a basic
scenario. We thus have to assume that muons experience a \emph{well
defined} internal field $H_\mu$ resulting from the well established
ordered spin ice structure on the length scale $l_C$. In order to
suppress oscillations of the muon relaxation, $H_\mu$ has to
\emph{fully} fluctuate in direction so that the mean internal field
vector at the muon site cancels out. For a fluctuation rate $\nu
\gtrsim \gamma_\mu H_\mu$, one then gets an exponential-like decay
of $P(t)$ (see for instance a comprehensive calculation for this
case in~\cite{keren04b}). This is understandable here since the
ordered spin ice state is six fold degenerate, the degrees of
freedom being the choice of the two in and two out spins out of the
four spins of one tetrahedron or equivalently of the direction of
the resulting moment among one of the six (100) type directions. We
thus propose that whole spin clusters, \emph{i.e.} domains of
typical size $l_C$ where the tetrahedrons are all in the same
configuration, fluctuate in between the six allowed configurations.
Due to the high symmetry of the proposed ground state, the mean
vector field at the muon site effectively vanishes. The correlation
length $l_C$ measured by neutron experiments~\cite{Mirebeau05}
remains limited to 18~nm well below $T_C$. It is therefore
consistent with the proposed cluster scenario. The small fluctuation
rate $\nu$ we found also supports cluster fluctuations rather than
single spin paramagnetic ones.
In the proposed picture, the "ferromagnetic" transition corresponds
to the freezing of the spin correlations on a large but finite
length scale $l_C$ which defines the extent of "long range" order.
This order is dynamical in the sense that the ordered clusters
undergo global rotations while the spin spin correlations are
preserved.

\begin{figure}
\includegraphics[scale=0.8]{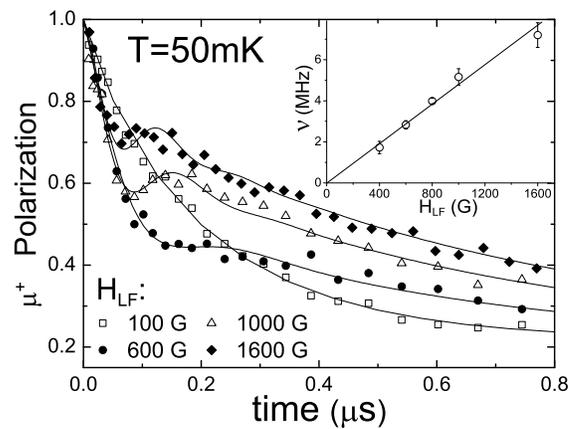}
\caption{\label{oscillations} Field dependence of the polarization
relaxation for not too large fields. The solid lines are
phenomenological fits to the sum of a damped cosine and a stretched
exponential functions used to track the early time oscillation
frequency $\nu$ plotted in the inset as a function of $H_{LF}$.}
\end{figure}

In our $\mu$SR study, the "long range ordered" ground state of
Tb$_2$Sn$_2$O$_7$ is only clearly evidenced in the detailed analysis
of magnetic field effects on the $\mu^+$ relaxation. As shown in
Fig.~\ref{oscillations}, for intermediate values of the applied
longitudinal fields $H_{LF}$, smaller than the ones used in
Fig.~\ref{decouplage}, the relaxation is no more exponential but
exhibits a strongly damped oscillation at early times. The frequency
of this oscillation is plotted in the inset of
Fig.~\ref{oscillations}. It scales linearly with the applied field
according to $2 \pi \nu / \gamma_\mu = A \times H_{LF}$ with $A
\simeq 0.37$. A common situation where oscillations result from an
applied longitudinal field appears in disordered magnets either
above or below their transition temperature, when the field at the
muon site $H_{\mu}$ is comparable to the applied field $H_{LF}$. The
muon spin then precesses around the total field
$\mathbf{H_{LF}+H_{\mu}}$ and due to the disordered nature of
$H_{\mu}$, either dynamical or static, the resulting polarization
shows a highly damped oscillation at a frequency $\gamma_\mu H_{LF}/
2 \pi$\cite{hayano79,keren02}. The small value of $A$ rules out such
an explanation here and the oscillations are more likely intrinsic,
\emph{i.e.} they arise from "long range order" in agreement with the
proposed cluster scenario. $A$ should then rather be associated with
a magnetic susceptibility  resulting from progressive spin canting
along the applied field. The effect of the external field is to
break the high symmetry of the ground state. The mean vector field
at the muon site is no more zero and oscillations are observed.


\begin{figure}
\includegraphics[scale=0.8]{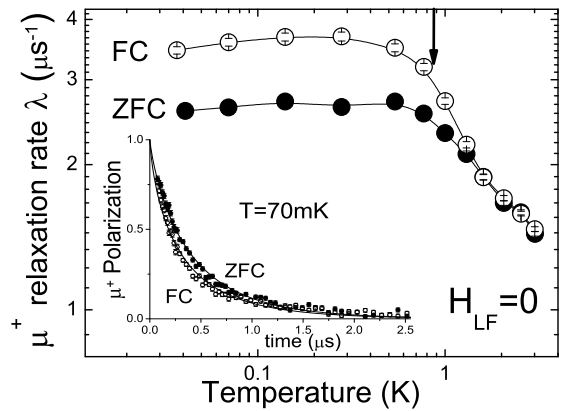}
\caption{\label{FCZFC} Muon relaxation rate $\lambda$ deduced from
fits with Eq.~\ref{eqstretched} of the zero field polarization
versus time data measured after cooling down the sample below 2~K in
zero external field (ZFC) or with $H_{LF}=800$~G (FC). The inset
shows the raw zero field polarization at base temperature for the
two cooling procedures.}
\end{figure}

After applying increasing longitudinal fields at base temperature as described above, we
surprisingly  did not recover the same polarization
decay in zero field. Namely, the relaxation rate was slightly higher
after the field experiment than before, although the polarization
versus time curves are qualitatively similar (see inset in Fig.~\ref{FCZFC}).
Relatively strong applied field ($H_{LF}
\gtrsim 800$~Oe) are likely to drive the system in a different spin configuration
resulting in a higher field $H_{\mu}+\Delta H_{\mu}$ at the muon
site.
Considering the small magnitude of $\Delta H_{\mu}$  (a fit with Eq.~\ref{eqstretched} gives $\Delta H_{\mu}/H_{\mu} \simeq 15\%$ at
$35$~mK) one can assume that only minor modifications of the
overall ordered spin ice structure subsist after application and
removal of the magnetic field. We checked that the
effect of an external magnetic field as presented in
Fig.~\ref{decouplage} and \ref{oscillations} hardly depends on the initial zero field state.
Thus, despite a slight change in the spin configurations, the ground state remains dynamical.
Such an hysteresis effect is a signature of frozen spin correlations as we expect below $T_C$. Field history dependence have for instance been reported in the spin ice Ho$_2$Ti$_2$O$_7$ compound in neutron~\cite{Harris97} and $\mu$SR~\cite{Harris98} experiments.
 We further investigated this effect by measuring the muon
relaxation rate as a function of temperature from 70~mK up to 3~K in
zero external field after (i) cooling down the sample from 3~K in
zero external field and (ii) cooling down the sample with
$H_{LF}$=800~G and then removing the external field at base
temperature. The results are plotted in Fig.~\ref{FCZFC}.
Interestingly the hysteresis starts at a somewhat higher temperature
$T \simeq 1.3$~K than the actual transition temperature defined by
the peak in the heat capacity measurements at $T_C=0.87$~K. This
perfectly agrees with neutron measurements where the magnitude of
the frozen Tb moment was shown to vanish only above 1.3~K. These
results point at a rather broad transition regime.

 To summarize, we
have evidenced that a dynamical regime survives in Tb$_2$Sn$_2$O$_7$
far below its ferromagnetic transition. This compound therefore
joins the recently exhibited class of pyrochlores where fluctuations
and long range order are simultaneously observed, namely
Gd$_2$Ti$_2$O$_7$, Gd$_2$Sn$_2$O$_7$ and Er$_2$Ti$_2$O$_7$. Our
results for Tb$_2$Sn$_2$O$_7$ are however markedly different from
the Gd$_2$Ti$_2$O$_7$ case, where fluctuations were observed in
addition to oscillations of the muon relaxation. Here the
exponential decay of the muon polarization  cannot be ascribed to
fluctuations around an overall long range order but results from
complete fluctuation of the local field. To get a consistent picture
with neutron results~\cite{Mirebeau04}, we proposed that whole
clusters of well ordered spins fluctuate in between the six
degenerate configurations allowed in the "ordered spin ice"
structure. Application of an external longitudinal field breaks the
symmetry of the configuration sub-space and oscillations can be
detected. Frozen correlations are also evidenced by a field history
dependence of the zero field muon relaxation below $\simeq T_C$.

The large number of spins involved in the coherent rotation of a
cluster demand a unique configuration space with vanishingly small
energy barriers in between the six fold degenerate configurations of
the ground state in order to allow for low energy collective
excitations. This is rather counterintuitive with respect to the
Ising type anisotropy which stabilizes the local spin ice order.
Contrary to spin ice pyrochlores, the "ordered spin ice" structure
may not be dominantly driven by single ion anisotropy and the latter
may not be too strong in Tb$_2$Sn$_2$O$_7$. It is also noticeable
that, despite the ferromagnetic transition of Tb$_2$Sn$_2$O$_7$, the
muon relaxation is strikingly similar to the well studied parent
compound Tb$_2$Ti$_2$O$_7$.
One may assume that the dynamics observed here
 is not different in nature from the collective excitations of the disordered Tb$_2$Ti$_2$O$_7$
 magnet~\cite{Gardner99,keren04}.
 Understanding the exotic dynamical ground state of Tb$_2$Sn$_2$O$_7$ where correlations are well defined may in turn bring new insight
 to the puzzling and more complex case of the Ti counterpart.

This research project 
has been supported by the European Commission under Framework
Programme 6 through the Key Action: Strengthening the European
Research Area, Research Infrastructures. Contract no:
RII3-CT-2003-505925.



\begin{thebibliography}{19}
\expandafter\ifx\csname
natexlab\endcsname\relax\def\natexlab#1{#1}\fi
\expandafter\ifx\csname bibnamefont\endcsname\relax
  \def\bibnamefont#1{#1}\fi
\expandafter\ifx\csname bibfnamefont\endcsname\relax
  \def\bibfnamefont#1{#1}\fi
\expandafter\ifx\csname citenamefont\endcsname\relax
  \def\citenamefont#1{#1}\fi
\expandafter\ifx\csname url\endcsname\relax
  \def\url#1{\texttt{#1}}\fi
\expandafter\ifx\csname urlprefix\endcsname\relax\def\urlprefix{URL
}\fi \providecommand{\bibinfo}[2]{#2}
\providecommand{\eprint}[2][]{\url{#2}}

\bibitem[{\citenamefont{Gardner et~al.}(1999)\citenamefont{Gardner, Dunsiger,
  Gaulin, Gingras, Greedan, Kiefl, Lumsden, MacFarlane, Raju, Sonier
  et~al.}}]{Gardner99}
\bibinfo{author}{\bibfnamefont{J.~S.}~\bibnamefont{Gardner \emph{et~al}.}},
   \bibinfo{journal}{Phys. Rev. Lett.}
  \textbf{\bibinfo{volume}{82}}, \bibinfo{pages}{1012} (\bibinfo{year}{1999}).

\bibitem[{\citenamefont{Pentrenko et~al.}(2004)\citenamefont{Pentrenko, Lees,
  Balakrishnan, and Paul}}]{Pentrenko04}
\bibinfo{author}{\bibfnamefont{O.}~\bibnamefont{Pentrenko}},
  \bibinfo{author}{\bibfnamefont{M.}~\bibnamefont{Lees}},
  \bibinfo{author}{\bibfnamefont{G.}~\bibnamefont{Balakrishnan}},
  \bibnamefont{and} \bibinfo{author}{\bibfnamefont{D.~M.} \bibnamefont{Paul}},
  \bibinfo{journal}{Phys. Rev. B} \textbf{\bibinfo{volume}{70}},
  \bibinfo{pages}{012402} (\bibinfo{year}{2004}).

\bibitem[{\citenamefont{Mirebeau and Goncharenko}(2004)}]{Mirebeau04}
\bibinfo{author}{\bibfnamefont{I.}~\bibnamefont{Mirebeau}} \bibnamefont{and}
  \bibinfo{author}{\bibfnamefont{I.}~\bibnamefont{Goncharenko}},
  \bibinfo{journal}{Phys. Rev. Lett.} \textbf{\bibinfo{volume}{93}},
  \bibinfo{pages}{187204} (\bibinfo{year}{2004}).

\bibitem[{\citenamefont{Matsuhira et~al.}(2002)\citenamefont{Matsuhira,
  Hinatsu, Tenya, Amitsuka, and Sakakibara}}]{Matsuhira02}
\bibinfo{author}{\bibfnamefont{K.}~\bibnamefont{Matsuhira \emph{et~al}.}},
  \bibinfo{journal}{J. Phys. Soc. Jpn} \textbf{\bibinfo{volume}{71}},
  \bibinfo{pages}{1576} (\bibinfo{year}{2002}).

\bibitem[{\citenamefont{Mirebeau et~al.}(2005)\citenamefont{Mirebeau, Apetrei,
  Rodriguez-Carjaval, Bonville, Forget, Colson, Glazkov, Sanchez, Isnard, and
  Suard}}]{Mirebeau05}
\bibinfo{author}{\bibfnamefont{I.}~\bibnamefont{Mirebeau \emph{et~al}.}},
  \bibinfo{journal}{Phys. Rev. Lett.} \textbf{\bibinfo{volume}{94}},
  \bibinfo{pages}{246402} (\bibinfo{year}{2005}).

\bibitem[{\citenamefont{Snyder et~al.}(2001)\citenamefont{Snyder, Slusky, Cava, and Schiffer}}]{Snyder01}
\bibinfo{author}{\bibfnamefont{J.}~\bibnamefont{Snyder}},
\bibinfo{author}{\bibfnamefont{J.~S.}~\bibnamefont{Slusky}},
\bibinfo{author}{\bibfnamefont{R.~J.}~\bibnamefont{Cava}},
\bibnamefont{and}
\bibinfo{author}{\bibfnamefont{P.}~\bibnamefont{Schiffer}},
  \bibinfo{journal}{Nature} \textbf{\bibinfo{volume}{413}},
  \bibinfo{pages}{48} (\bibinfo{year}{2001}).
\bibinfo{author}{\bibfnamefont{S.~T.}~\bibnamefont{Bramwell}},
\bibnamefont{and}
\bibinfo{author}{\bibfnamefont{M~.J~.P.}~\bibnamefont{Gingras}},
  \bibinfo{journal}{Science} \textbf{\bibinfo{volume}{294}},
  \bibinfo{pages}{1495} (\bibinfo{year}{2001}).


\bibitem[{\citenamefont{Uemura et~al.}(1994)\citenamefont{Uemura, Keren,
  Kojima, Le, Luke, Wu, Ajiro, Asano, Kuriyama, Mekata et~al.}}]{Uemura94}
\bibinfo{author}{\bibfnamefont{Y.}~\bibnamefont{Uemura \emph{et~al}.}},
  \bibinfo{journal}{Phys. Rev. Lett.}
  \textbf{\bibinfo{volume}{73}}, \bibinfo{pages}{3306} (\bibinfo{year}{1994}).

\bibitem[{\citenamefont{Dunsiger et~al.}(1996)\citenamefont{Dunsiger, Kiefl,
  Chow, Gaulin, Gingras, Greedan, Keren, Kojima, Luke, MacFarlane
  et~al.}}]{Dunsiger96}
\bibinfo{author}{\bibfnamefont{S.}~\bibnamefont{Dunsiger \emph{et~al}.}},
 \bibinfo{journal}{Phys. Rev. B}
  \textbf{\bibinfo{volume}{54}}, \bibinfo{pages}{9019} (\bibinfo{year}{1996}).

\bibitem[{\citenamefont{Bert et~al.}(2005)\citenamefont{Bert, Bono, Mendels,
  Ladieu, Duc, Trombe, and Millet}}]{Bert05}
\bibinfo{author}{\bibfnamefont{F.}~\bibnamefont{Bert \emph{et~al}.}},
  \bibinfo{journal}{Phys. Rev. Lett.} \textbf{\bibinfo{volume}{95}},
  \bibinfo{pages}{087203} (\bibinfo{year}{2005}).

\bibitem[{\citenamefont{Bono et~al.}(2004)\citenamefont{Bono, Mendels, Collin,
  Blanchard, Bert, Amato, Baines, and Hillier}}]{Bono04b}
\bibinfo{author}{\bibfnamefont{D.}~\bibnamefont{Bono \emph{et~al}.}},
  \bibinfo{journal}{Phys. Rev. Lett.} \textbf{\bibinfo{volume}{93}},
  \bibinfo{pages}{187201} (\bibinfo{year}{2004}).

\bibitem[{\citenamefont{Bertin et~al.}(2002)\citenamefont{Bertin, Bonville,
  Bouchaud, Hodges, Sanchez, and Vulliet}}]{Bertin02}
\bibinfo{author}{\bibfnamefont{E.}~\bibnamefont{Bertin \emph{et~al}.}},
  \bibinfo{journal}{Eur. Phys. J. B} \textbf{\bibinfo{volume}{27}},
  \bibinfo{pages}{347} (\bibinfo{year}{2002}).

\bibitem[{\citenamefont{Yaouanc et~al.}(2005)\citenamefont{Yaouanc,
  de~R{\'e}otier, Glazkov, Marin, Bonville, Hodges, Gubbens, Sakarya, and
  Baines}}]{Yaouanc05}
\bibinfo{author}{\bibfnamefont{A.}~\bibnamefont{Yaouanc \emph{et~al}.}},
  \bibinfo{journal}{Phys. Rev. Lett.} \textbf{\bibinfo{volume}{95}},
  \bibinfo{pages}{047203} (\bibinfo{year}{2005}).

\bibitem[{\citenamefont{Lago et~al.}(2005)\citenamefont{Lago, Lancaster,
  Blundell, Bramwell, Pratt, Shirai, and Baines}}]{Lago05}
\bibinfo{author}{\bibfnamefont{J.}~\bibnamefont{Lago \emph{et~al}.}},
  \bibinfo{journal}{J. Phys.: Condens. Matter} \textbf{\bibinfo{volume}{17}},
  \bibinfo{pages}{979} (\bibinfo{year}{2005}).

\bibitem[{\citenamefont{Keren et~al.}(2004)\citenamefont{Keren, Gardner,
  Ehlers, Fukaya, Segal, and Uemura}}]{keren04}
\bibinfo{author}{\bibfnamefont{A.}~\bibnamefont{Keren \emph{et~al}.}},
  \bibinfo{journal}{Phys. Rev. Lett.} \textbf{\bibinfo{volume}{92}},
  \bibinfo{pages}{107204} (\bibinfo{year}{2004}).

\bibitem[{\citenamefont{Keren}(2004)}]{keren04b}
\bibinfo{author}{\bibfnamefont{A.}~\bibnamefont{Keren}}, \bibinfo{journal}{J.
  Phys.: Condens. Matter} \textbf{\bibinfo{volume}{16}}, \bibinfo{pages}{S4603}
  (\bibinfo{year}{2004}).

\bibitem[{\citenamefont{Hayano et~al.}(1979)\citenamefont{Hayano, Uemura,
  Imazato, Nishida, Yamazaki, and Hubo}}]{hayano79}
\bibinfo{author}{\bibfnamefont{R.}~\bibnamefont{Hayano \emph{et~al}.}},
  \bibinfo{journal}{Phys. Rev. B} \textbf{\bibinfo{volume}{20}},
  \bibinfo{pages}{850} (\bibinfo{year}{1979}).

\bibitem[{\citenamefont{Gingras et~al.}(2000)\citenamefont{Gingras, Hertog,
  Faucher, Gardner, Dunsiger, Chang, Gaulin, Raju, and Greedan}}]{gingras00}
\bibinfo{author}{\bibfnamefont{M.}~\bibnamefont{Gingras \emph{et~al}.}},
  \bibinfo{journal}{Phys. Rev. B} \textbf{\bibinfo{volume}{62}},
  \bibinfo{pages}{6496} (\bibinfo{year}{2000}).

\bibitem[{\citenamefont{Keren et~al.}(2002)\citenamefont{Keren, Gulener,
  Campbell, Bazalitsky, and Amato}}]{keren02}
\bibinfo{author}{\bibfnamefont{A.}~\bibnamefont{Keren}},
  \bibinfo{author}{\bibfnamefont{F.}~\bibnamefont{Gulener}},
  \bibinfo{author}{\bibfnamefont{I.}~\bibnamefont{Campbell}},
  \bibinfo{author}{\bibfnamefont{G.}~\bibnamefont{Bazalitsky}},
  \bibnamefont{and} \bibinfo{author}{\bibfnamefont{A.}~\bibnamefont{Amato}},
  \bibinfo{journal}{Phys. Rev. Lett.} \textbf{\bibinfo{volume}{89}},
  \bibinfo{pages}{107201} (\bibinfo{year}{2002}).

\bibitem[{\citenamefont{Luo et~al.}(2001)\citenamefont{Luo, Hess, and
  Corruccini}}]{Luo01}
\bibinfo{author}{\bibfnamefont{G.}~\bibnamefont{Luo}},
  \bibinfo{author}{\bibfnamefont{S.}~\bibnamefont{Hess}}, \bibnamefont{and}
  \bibinfo{author}{\bibfnamefont{L.}~\bibnamefont{Corruccini}},
  \bibinfo{journal}{Phys. Lett. A} \textbf{\bibinfo{volume}{291}},
  \bibinfo{pages}{306} (\bibinfo{year}{2001}).



\bibitem[{\citenamefont{Harris et~al.}(1997)\citenamefont{Harris, Bramwell, McMorrow,
Zeiske, and Godfrey}}]{Harris97}
\bibinfo{author}{\bibfnamefont{M.~J.}~\bibnamefont{Harris}},
  \bibinfo{author}{\bibfnamefont{S.~T.}~\bibnamefont{Bramwell}},
  \bibinfo{author}{\bibfnamefont{D.~F.}~\bibnamefont{McMorrow}},
  \bibinfo{author}{\bibfnamefont{T.}~\bibnamefont{Zeiske}},
  \bibnamefont{and}
  \bibinfo{author}{\bibfnamefont{K.~W.}~\bibnamefont{Godfrey}},
  \bibinfo{journal}{Phys. Rev. Lett.} \textbf{\bibinfo{volume}{79}},
  \bibinfo{pages}{2554} (\bibinfo{year}{1997}).

\bibitem[{\citenamefont{Harris et~al.}(1998)\citenamefont{Harris, Bramwell, Zeiske, McMorrow,
 and King}}]{Harris98}
\bibinfo{author}{\bibfnamefont{M.~J.}~\bibnamefont{Harris}},
  \bibinfo{author}{\bibfnamefont{S.~T.}~\bibnamefont{Bramwell}},
   \bibinfo{author}{\bibfnamefont{T.}~\bibnamefont{Zeiske}},
  \bibinfo{author}{\bibfnamefont{D.~F.}~\bibnamefont{McMorrow}},
  \bibnamefont{and}
  \bibinfo{author}{\bibfnamefont{P.~J.~C.}~\bibnamefont{King}},
  \bibinfo{journal}{J. Magn. Magn. Mater.} \textbf{\bibinfo{volume}{177}},
  \bibinfo{pages}{757} (\bibinfo{year}{1998}).

\end{thebibliography}

\end{document}